%% file: template.tex
\documentclass[journal]{vgtc}                     

\onlineid{1343}

\vgtccategory{Research}

\vgtcpapertype{application/design study}

\title{GeoExplainer: A Visual Analytics Framework for Spatial Modeling Contextualization and Report Generation}

\author{
    Fan~Lei, 
    Yuxin~Ma, \textit{Senior Member, IEEE},
    A.~Stewart~Fotheringham, 
    Elizabeth~A.~Mack, 
    Ziqi~Li, \\
    Mehak~Sachdeva,
    Sarah~Bardin,
    and Ross~Maciejewski, \textit{Senior Member, IEEE}}
\authorfooter{
\item
 F. Lei, A.S. Fotheringham, M. Sachdeva, S. Bardin, and R. Maciejewski are with Arizona State University. E-mail: \{flei5,sfotheri,msachde1,sfbardin,rmacieje\}@asu.edu.
\item
 Y. Ma is with Southern University of Science and Technology. E-mail: mayx@sustech.edu.cn. Y. Ma is the corresponding author.
\item
 E.A. Mack is with Michigan State University. E-mail: emack@msu.edu.
 \item
 Z. Li is with Florida State University. E-mail: liziqi1992@gmail.com.
}

\abstract{%
  Geographic regression models of various descriptions are often applied to identify patterns and anomalies in the determinants of spatially distributed observations. These types of analyses focus on answering why questions about underlying spatial phenomena, e.g., why is crime higher in this locale, why do children in one school district outperform those in another, etc.? Answers to these questions require explanations of the model structure, the choice of parameters, and contextualization of the findings with respect to their geographic context. This is particularly true for local forms of regression models which are focused on the role of locational context in determining human behavior. In this paper, we present GeoExplainer, a visual analytics framework designed to support analysts in creating explanative documentation that summarizes and contextualizes their spatial analyses. As analysts create their spatial models, our framework flags potential issues with model parameter selections, utilizes template-based text generation to summarize model outputs, and links with external knowledge repositories to provide annotations that help to explain the model results. As analysts explore the model results, all visualizations and annotations can be captured in an interactive report generation widget. We demonstrate our framework using a case study modeling the determinants of voting in the 2016 US Presidential Election.
}

\keywords{Spatial data analysis, local models, multiscale geographically weighted regression, model explanation, visual analytics}

\teaser{
  \centering
  \includegraphics[width=1.0\linewidth]{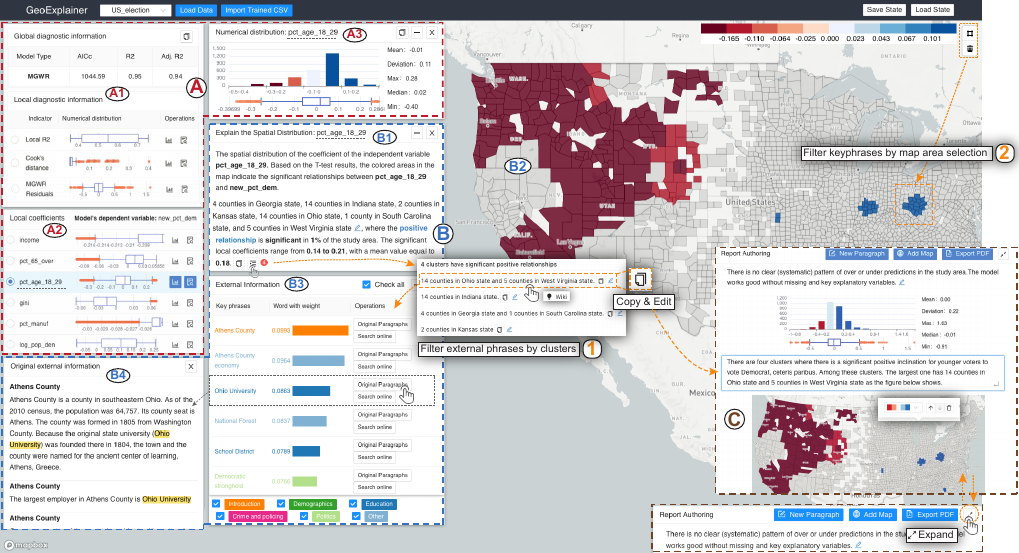}
  \caption{The model analysis and report authoring interface. The model validation module (A) contains: the model diagnostic panel (A1); a local coefficient list (A2), and; the numerical distribution panel (A3) showing the distribution of the selected coefficient of \textit{pct\_age\_18\_29}. The contextualized exploration module (B) contains: the local parameter explanation view (B1); the map view (B2) where a choropleth map presents the spatial distribution of a selected coefficient or diagnostic value; the external information module, including an external keyphrase list (B3) which retrieves the most important keyphrases from the external context related to the selected range of regions, and; a list of the original external content (B4) that consists of paragraphs of relevant external information related to a selected phrase. The report authoring module (C) supports synthesizing the findings into a report.}
\label{fig:teaser}
}

\graphicspath{{figs/}{figures/}{pictures/}{images/}{./}} 

\usepackage{tabu}                      
\usepackage{booktabs}                  
\usepackage{lipsum}                    
\usepackage{mwe}                       

\usepackage{mathptmx}                  
\usepackage{microtype}                 
\PassOptionsToPackage{warn}{textcomp}  
\usepackage{amsmath}
\usepackage{textcomp}
\usepackage{cite}                      
\usepackage{wrapfig}

\usepackage[table,xcdraw,svgnames]{xcolor}
\definecolor{TODOcolor}{HTML}{df65b0} 
\definecolor{AddContentcolor}{HTML}{0000CC}
\definecolor{patternColor}{HTML}{1C75BC}
\definecolor{resultColor}{HTML}{00A14B}

\newcommand{\NEW}[0]{}

\newcommand{\PHASEONE}[1]{\textcolor{patternColor}{#1}}
\newcommand{\PHASETWO}[1]{\textcolor{resultColor}{#1}}

\usepackage[normalem]{ulem} 

\usepackage{lscape}

\usepackage{inconsolata}

\renewcommand{\sout}[1]{}

\begin{document}



\maketitle
\input{content/1_introduction.tex}

\input{content/2_related_work.tex}
\input{content/3_overview.tex}

\input{content/4_framework_design.tex}

\input{content/5_evaluation.tex}
\input{content/6_conclusion.tex}

\input{content/supplemental_materials.tex}

\acknowledgments{
The authors wish to thank Qiang Xu for the engineering support in the framework development.
This work was supported in part by the National Science Foundation under Grant 1639227, the U.S. Department of Homeland Security under Award 2017-ST-061-QA0001, the National Natural Science Foundation of China (No. 62202217), Guangdong Basic and Applied Basic Research Foundation (No. 2023A1515012889), and Guangdong Talent Program (No. 2021QN02X794).
}


\bibliographystyle{abbrv-doi-hyperref}

\bibliography{template}

\appendix 
\onecolumn
\centering
\pagenumbering{arabic} 
\markboth{}{} 

\input{content/7_appendix}

\end{document}

%% file: content/1_introduction.tex
\section{Introduction}
\label{sec:introduction}


Geographic models~\cite{fotheringham2003geographically, fotheringham2017multiscale, oshan2019mgwr} are often used to answer \textit{why} questions regarding underlying spatial phenomena such as why do voter preferences generally exhibit strong spatial dependency. In order to answer such questions, researchers have developed sophisticated spatial modeling techniques and software, such as multiscale geographically weighted regression (MGWR)~\cite{fotheringham2017multiscale, oshan2019mgwr}, to identify the determinants of the spatial patterns of data we observe in both the human and natural environments. However, these models generate output that needs local contextual information to be interpreted properly. Without such information, advanced spatial modeling techniques do not gain their full potential. 
 While a variety of geovisual analytics systems~\cite{Guo2006, Eccles2008, Lundblad2013, Ferreira2013, Li2015, Latif2019} have been developed to explore spatial statistics with the help of interactive maps and narrative annotations, these applications tend to visualize only basic statistical results, and, to our knowledge, none of them explain spatial modeling results.
 
 
In this paper, we present GeoExplainer (Figure~\ref{fig:teaser}), a visual analytics framework designed to support spatial data modeling, analysis and reporting. Our approach is inspired by research in explainable machine learning~\cite{krause2017workflow, krause2014infuse} and a combination of narrative visualization~\cite{Segel2010} and storytelling techniques~\cite{kosara2013storytelling}. GeoExplainer provides multiple types of explanation support throughout the spatial analysis pipeline. In the model calibration stage, our framework interprets the functionality of the spatial model configuration and recommends parameter settings. Then, the framework summarizes and explains model outputs by adopting template-based text annotations, linking with external knowledge repositories to provide relevant contextual information. All visualizations, model results, and annotations can be captured in an interactive report authoring widget, enabling analysts to generate documentation that explains their spatial analyses. Our contributions include:
\begin{itemize}[leftmargin=*]

\item A novel narrative explanation workflow that supports a dynamic interplay between automatically generated text interpretations and relevant visualizations.

\item An automatic spatial cluster detection function that suggests interesting patterns and outliers from the local spatial modeling result.

\item The seamless integration of an interactive report authoring tool with the spatial analysis pipeline.

\item A web-based visual analytics framework with analytical process recording and sharing functions to facilitate collaboration.


\end{itemize}

%% file: content/2_related_work.tex
\section{Related Work}
\label{sec:related_work}

Our work focuses on facilitating spatial analysis by explaining the outputs of spatial modeling with contextual information through narrative visualization. In this section, we review related work on geographic analysis, narrative visualization, and model explainability.

\subsection{Geographic Analysis}

Various spatial prediction models~\cite{kanevski2009machine, li2011application, liess2012uncertainty, zhu2018spatial} and spatial data analysis tools~\cite{Takatsuka2002, Guo2006} have been developed to support geographic analysis. In this work, we focus on two local spatial models widely used in spatial analysis: Geographically Weighted Regression (GWR)~\cite{brunsdon1996geographically}, and its recent extension, Multiscale Geographically Weighted Regression (MGWR)~\cite{fotheringham2017multiscale}. \NEW{GWR extends the classical linear regression model~\cite{kramer1987spatial} by capturing spatial heterogeneity with influence spreading over the space in a constant scale. MGWR further improves GWR models where local influences are modeled in different spatial windows.}

Support for a variety of spatial models (including GWR and MGWR) is integrated into the most widely used geographic information systems (GIS), e.g. ArcGIS~\cite{scott2010spatial}, QGIS~\cite{gil2015space} and GeoDa Web~\cite{Li2015}, and embedded by libraries (e.g., PySAL~\cite{rey2010pysal}) in computational notebook environments such as Jupyter Notebook~\cite{ozturk2016geonotebook, oshan2019mgwr, rey2021pysal} or R Markdown~\cite{lovelace2019geocomputation}.  
However, these systems and notebooks do not support integrated external knowledge sources to help contextualize models. \NEW{They also require analysts to judiciously choose the proper tools at every stage of the analytical pipeline to build a well-trained model. This process necessitates a solid foundation in geographical concepts and can be time-consuming, even for seasoned analysts.} Our goal is to enhance the spatial modeling process through a deliberately designed user-friendly workflow and automatically generated explanatory narratives guided by domain experts to promote the understanding of local contextual information from models. Our work focuses on improving model explainability, enabling interactive report authoring, and supporting contextualization through narrative generation. This is important in local models such as GWR and MGWR because the main output from such models is a set of local parameter estimates from each process being modeled, and the spatial variation in these estimates needs to be explained in terms of the contextualized environment of each location by providing additional contextual information for users.


\subsection{Narratives and Annotations} 
The design space of narrative visualization~\cite{Segel2010, Hullman2011, Tong2018} has been widely explored in the visualization community and a diverse set of storytelling and annotation methods have been developed to reveal observations in data and to convey key messages to an audience. 
Such narrative visualization techniques have been adopted by a variety of geovisual analytics applications~\cite{Eccles2008, Lundblad2013, Satyanarayan2014} that have integrated story authoring tools with spatial data visualizations. 
For example, Bespoke Map~\cite{Brock2017} discusses design implications for map customization tools. NewsViews~\cite{Gao2014} generates interactive maps with narrative annotations automatically from a given news article. Latif and Beck~\cite{Latif2019} introduce a bivariate map design that integrates template-based text annotations, and they later extend their work to investigate the interplay of text and visualization in geographic storytelling~\cite{Latif2021}. 

Many of these systems utilize annotations to enhance the visual narratives, and Kosara and Mackinlay~\cite{kosara2013storytelling} emphasize the importance of annotations for facilitating storytelling in visualization. 
Recent annotation work~\cite{chen2010touch2annotate, chen2010click2annotate, ren2017chartaccent, Satyanarayan2014} has focused on the semi-automatic creation of presentation-like storytelling visualizations and explored mechanisms for automatically generating annotations by integrating deep-learning feature extraction techniques with a natural language generation process~\cite{Lai2020, Liu2020}. 
Annotations in such systems are used to describe salient patterns and facilitate storytelling~\cite{Cui2019a, Srinivasan2019, Chen2020, choudhry2020once}. 

In addition to using annotations to explain data features, other works have also explored mechanisms for supplementing narratives from external data corpora. For example, Contextifier~\cite{Hullman2013} and Causeworks~\cite{choudhry2020once} leverage relevant external information from news articles or Wikipedia to generate text annotations. Such external information provides audiences with additional context to facilitate the analysis. 

While advanced storytelling techniques are being readily adopted to develop narrative reports for geovisualization, the focus has been on supporting simple statistical analysis (e.g., find the maximum, minimum). However, modeling and analysis processes have increased well beyond descriptive statistics, and there is a growing need to expand the narrative generation process to support more sophisticated spatial models, such as MGWR~\cite{fotheringham2017multiscale, oshan2019mgwr}. Our work builds upon automatic narrative generation and interactive annotation research to support explanations, annotations, and contextualization of spatial modeling. We provide templates for text generation for model explainability and develop interactive authoring tools for report generation.


\subsection{Model Explainability}
Narrative visualization is directly related to the concept of explainability, where the visualization authors seek to couple images and text to explain an underlying data analysis. With respect to model explainability, the visual analytics community has developed a variety of systems to support the interactive explanation of machine learning models (e.g.~\cite{bertini2009surveying, liu2017towards, lu2017recent, lu2017state}). Several model-independent approaches, e.g. EnsembleMatrix~\cite{talbot2009ensemblematrix} and RuleMatrix~\cite{ming2018rulematrix}, focus on the classifier's input-output behaviors to provide insight into the model classification results. EnsembleMatrix provides a visual summary of the model outputs, RuleMatrix uses a matrix-based visualization to explain classification results, and Prospector~\cite{krause2016interacting} explores the relationship between feature values and predictions by using partial dependence diagnostics. 


Most closely related to our work are the techniques that explain models from feature-level observation. Mühlbacher and Piringer~\cite{Muhlbacher2013} facilitated feature selection and optimization in regression models by partitioning the feature space into disjoint regions for visualization. 
Sedlmair et al.~\cite{Sedlmair2014} proposed an abstract conceptual framework to discuss the visual parameter space analysis problems independent of the application domain. 
Goodwin et al.~\cite{Goodwin2016} extended visual parameter space analysis to the spatial domain, enabling the exploration of correlations between multiple variables that vary geographically at different spatial scales. While these techniques focus on explanations for domain experts, our work is designed to support explanations to experts and support their use of external information to contextualize these relationships and \NEW{communicate \sout{explain}} their findings to a general audience. Our choice of text templates, as opposed to large language models, for narrative generation is to support control for reliability and reproducibility. All text generated must have a verifiable source and the generation of text for the analysis should always return the same results to ensure that the resulting analyses are not subject to misinformation.

%% file: content/3_overview.tex
\section{Design Overview}
\label{sec:design_overview}



Our goal is to implement an interactive visual analytical framework with \NEW{narrative contextualization techniques and} report authoring to support the spatial analysis pipeline.

\vspace{1.4mm} 
\noindent \textbf{\NEW{Design study with domain experts.}}\; 
\NEW{An iterative design process was employed to develop GeoExplainer in collaboration with two domain experts in quantitative geography. The first possesses over thirty years of experience and is renowned in spatial modeling and local statistical analysis. The second expert has over fifteen years of experience in geographic statistical analysis and applied geographic science. Our two-phase design process began with identifying spatial data analysis challenges and establishing initial design tasks through several one-on-one semi-structured interviews and group discussions, leading to the development of a conceptual demo. The second phase involved refining these design tasks and improving the framework's functionality and visual design through bi-weekly meetings for a year until all issues were resolved. This led to the identification of five design challenges (C1 - C5) and the formation of a set of analytical tasks (T1 - T5), with C1 demonstrating an architecture-level challenge and C2 to C5 pertain to individual stages of the spatial data analysis pipeline.}


\vspace{1.4mm} 
\noindent \textbf{C1: Flexible analytical pipeline.}\;
Although comprehensive spatial analysis frameworks (e.g., ArcGIS~\cite{scott2010spatial}, GeoDa~\cite{anselin2010geoda},  PySAL~\cite{rey2010pysal}) support sophisticated spatial modeling, \NEW{they need analysts to select a proper combination among the models and analytical tools to correctly perform the analysis. The analysts have to understand the key concepts of geographical and statistical to perform their own ESDA process~\cite{StewartFotheringham2021}.}
The challenge is making spatial modeling more accessible with a \NEW{\sout{unified}}framework \NEW{integrating a standard ESDA pipeline with dynamic interpretations to support the understanding and tools to enhance the modeling result communication. And this unified framework} does not require complex installation\NEW{, configuration,} and coding.
Design tasks for this challenge focus on implementing a framework to support a unified spatial analytical pipeline that is accessible to \NEW{\sout{a broad audience}}\NEW{the analysts.}

\begin{itemize}[leftmargin=*]
    \item \textbf{T1.1:} Develop an analytical pipeline that unifies ESDA, narrative explanations, and report authoring functionalities.
    \vspace{-0.6mm}
    \item \textbf{T1.2:} Support coordinated multiple views, each of which can populate information into the report authoring mechanics.
    \vspace{-0.6mm}
    \item \textbf{T1.3:} Instantiate the framework within modern web browsers to enable broader usage.
\end{itemize}

\noindent \textbf{C2: Model configuration and validation support.}\;
Configuring a modern spatial model, such as MGWR, with properly selected parameters, and validating performance features are fundamental prerequisites for the quality of the model output\cite{pace1997using, shabrina2021short}. However, strong prerequisite knowledge is needed to utilize relevant geostatistical tools and understand their results~\cite{oshan2019mgwr}. The challenge is to develop a framework that can guide \NEW{\sout{novices and experts}}\NEW{analysts} to appropriate model configurations and support the model validation process. Design tasks include:

\begin{itemize}[leftmargin=*]
    \item \textbf{T2.1:} Support the interactive selection of dependent variables and model parameters.
    \vspace{-0.6mm}
    \item \textbf{T2.2:} Generate interpretable parameter recommendations.
    \vspace{-0.6mm}
    \item \textbf{T2.3:} Visualize the model diagnostics and parameter estimates associated with each covariate.
\end{itemize}

\noindent \textbf{C3: Contextualizing model outputs.}\;
Once a spatial model is fit to the data, the parameter estimates and diagnostics are often produced in the form of data tables (e.g., comma-separated values files, JSON files, etc.)~\cite{oshan2019mgwr}. These numerical and statistical records in the table contain heterogeneous local contextual information, which is hard to understand without background information~\cite{oshan2019mgwr, StewartFotheringham2021}. Most extant tools only use maps to visualize the spatial surfaces that result from local modeling without the ability to bring in further contextual information. The challenge is how to automatically provide relevant contextual information to support explanations in the spatial analysis pipeline. Design tasks include:

\begin{itemize}[leftmargin=*]
    
    \item \textbf{T3.1:} Contextualize the model results.
    \vspace{-0.6mm}
    \item \textbf{T3.2:} Automatically identify interesting geographical patterns.
    \vspace{-0.6mm}
    \item \textbf{T3.3:} Explain the local parameter estimates in relation to the identified spatial patterns.
    \vspace{-0.6mm}
    \item \textbf{T3.4:} Link external contextual information associated with the identified spatial patterns or analyst selections.
\end{itemize}

\noindent \textbf{C4: Communicating findings.}\;
The goal of spatial analysis is to answer questions about spatial phenomena and communicate the findings to a broad audience~\cite{Chen2020}. However, the multifaceted analysis outcomes from advanced analytical tools are often difficult to interpret for a general audience (C3). The challenge is how to support the analysts in communicating the outputs of their spatial models. Design tasks for this challenge focus on storytelling and report authoring.

\begin{itemize}[leftmargin=*]
    \item \textbf{T4.1:} Highlight, annotate, and organize the analysts' findings in the form of text paragraphs and images.
    \vspace{-0.6mm}
    \item \textbf{T4.2:} Transform the findings into reports for a broad audience.
\end{itemize}

\noindent \textbf{C5: Collaboration and replicability.}\;
Analysts often need to save, share, or revert their analytical state during the model development process. The challenge is to develop a \NEW{asynchronized} spatial analysis framework that enables analysts to save\NEW{, retrieve,} and share their configured model, explanations, and mid-state analytics workflow~\cite{fotheringham2008sage}. Design tasks include:

\begin{itemize}[leftmargin=*]
    \item \textbf{T5.1:} Save the analytical state at any point during the pipeline, including the configured model, explored information, and the report.
    
    \vspace{-0.6mm}
    \item \textbf{T5.2:} \NEW{Retrieve} the stored state and continue the analysis.
    \vspace{-0.6mm}
    \item \textbf{T5.3:} Share the stored state with other analysts.
\end{itemize}

%% file: content/4_framework_design.tex
\section{GeoExplainer}
\label{sec:framework}

\begin{figure*}[t!]
	\centering	
	\includegraphics[width=2.00\columnwidth]{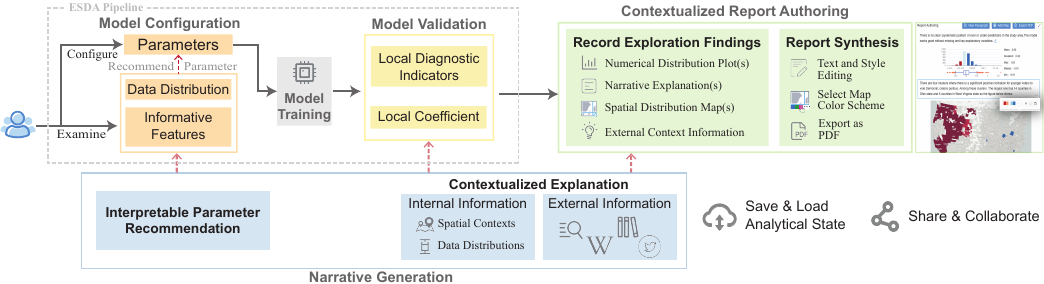}
	\caption{The GeoExplainer pipeline is designed to support the exploratory spatial data analysis (ESDA) workflow with Narrative Generation and Contextualized Report Authoring modules. The analytical state recording functions support \NEW{asynchronous analysis} and collaboration.}
	\label{fig:overall_pipeline}
	\vspace{-4mm}
\end{figure*}

Based on the design overview, we developed a visual analytics framework, GeoExplainer, to support spatial analysis across model creation, validation, contextualization, and reporting.

\begin{figure*}[t!]
	\centering	
	\includegraphics[width=2.00\columnwidth]{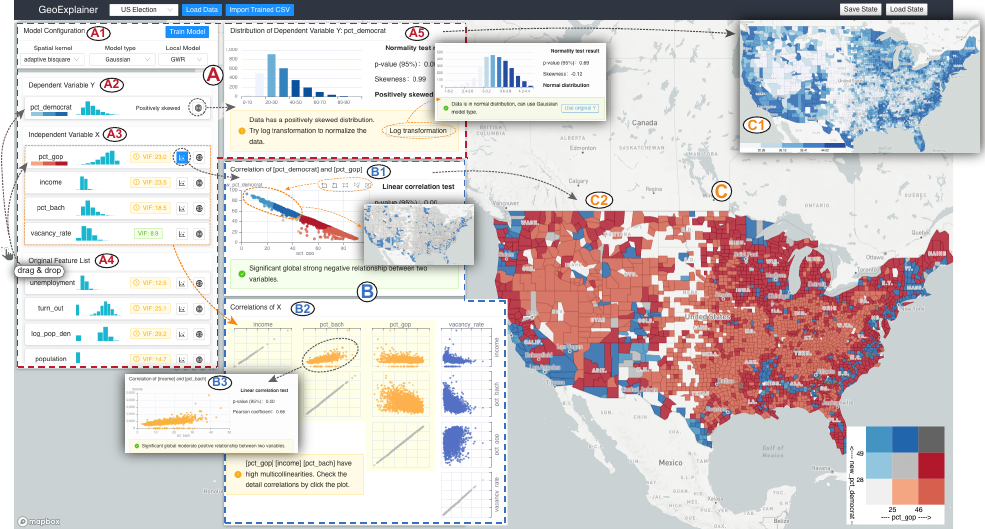}
	\caption{The model configuration interface. In the feature view (A), the \textit{model configuration panel} (A1) provides the configuration properties. The \textit{dependent variable selection panel} (A2) and \textit{independent variable selection panel} (A3) use the features that are dragged from the \textit{feature list} (A4), where each row is associated with a feature in the dataset with the marginal data distribution of the feature depicted as a histogram. The \textit{detailed feature analysis view} (A5) displays the histogram distribution and narrative information suggesting appropriate transformations to apply to the dependent variable. (B) The correlation view (B1) shows the correlation between the dependent variable and a selected independent variable. The \textit{scatterplot matrix} (B2) shows the correlations among multiple selected independent variables. The detailed distribution information with Pearson coefficients and text annotations (B3) is populated by clicking on the scatterplots in the matrix. (C) The map populates a \textit{univariable choropleth map} (C1) to show the spatial distribution of a selected variable. The map can also update a \textit{bivariate choropleth map} (C2) to display the spatial distributions and the degree of correlation between the dependent variable and a selected independent variable.}
	\label{fig:module_configuration}
	\vspace{-4mm}
\end{figure*}

\vspace{1.4mm} 
\noindent \textbf{Framework Pipeline:} The GeoExplainer framework has a unified pipeline (Figure~\ref{fig:overall_pipeline}) that enhances the exploratory spatial data analysis (ESDA) pipeline by adding contextualized interpretations and report synthesis functions (\textbf{T1.1}). The GeoExplainer pipeline is composed of five functional units: \emph{model configuration} (Sect. 4.1), \emph{model validation} (Sect. 4.2), \emph{narrative generation} (Sect. 4.3), \emph{contextualized report authoring} (Sect. 4.4), and \emph{analytical state recording} (Sect. 4.5). These functional units are spread across two coordinated multiple view interfaces. Figure~\ref{fig:module_configuration} illustrates the model configuration interface, and Figure~\ref{fig:teaser} shows the model validation and contextualized report authoring interface. Underlying both of these are narrative generation and analytical state recording tools. The narrative generation tools assist the audience in accessing the information of each module in the ESDA workflow, while the state recording module further supports analysts in saving, loading, and sharing analytical states throughout the pipeline.
In each functional unit, we link the dynamic graphics and narrative components with coordinated multiple views (\textbf{T1.2}) to support ESDA~\cite{anselin1996interactive, anselin2010geoda, Li2015}.

\vspace{1.4mm} \noindent \textbf{Web-based, Open-sourced Implementation:}\;
Analysts only need a web browser to utilize our framework (\textbf{T1.3}) as GeoExplainer is built upon a client-server architecture with a RESTful API between the front-end interface and back-end services. The back-end server is implemented with Python Flask, while the front-end interface uses React~\cite{React2022}, Mapbox GL JS~\cite{kastanakis2016mapbox}, and D3.js~\cite{bostock2011d3}.


The visual design and interactions are robust to various types of spatially referenced data and external knowledge sources for contextualization. Our target audience is \NEW{\sout{includes both experts and novices}}\NEW{spatial data analysts}, as our goal is to \NEW{help them save time and effort in spatial analysis while supporting their improved understanding and communication of the model results.}\sout{support spatial analysis for a broad range of analysts} For demonstration purposes, we focus on spatial analyses using \textbf{(MGWR)}, and a discussion on the generalization of this work is provided in Sect.6.

\subsection{Model Configuration}

The model configuration stage acts as an entry point for analysts to inspect data distributions, identify informative features, and find optimal model parameters before a spatial model is trained. Illustrated in Figure~\ref{fig:module_configuration}, the model configuration interface consists of three coordinated views: the \textbf{feature view}, the \textbf{correlation view}, and the \textbf{map view}.


\vspace{1.4mm} \noindent \textbf{Feature View:} The feature view, Figure~\ref{fig:module_configuration} (A), provides a summary of data distributions and evaluation results for features as well as the model parameter settings. It consists of five panels (A1-A5): the model configuration panel (A1), the dependent variable selection panel (A2), the independent variables selection panel (A3), the feature list (A4), and the detailed feature analysis view (A5).

\vspace{1.15mm}
\noindent In the \emph{model configuration panel}, Figure~\ref{fig:module_configuration} (A1), analysts select the configuration properties for MGWR, an adaptive bisquare spatial kernel is the default setting. For a detailed explanation of MGWR parameters, we refer the reader to the MGWR 2.2 Manual~\cite{oshan2019mgwr}. 

\vspace{1.15mm}
\noindent In the \emph{feature list}, Figure~\ref{fig:module_configuration} (A4), each row depicts the marginal data distribution of a feature as a histogram.
The features can be dragged into the \emph{dependent variable selection panel} (A2) or the \emph{independent variable selection panel} (A3) to configure the model.
An analyst starts the exploration of feature relationships by choosing a dependent variable. Then, our framework automatically performs multicollinearity detection using the variance inflation factor(VIF) to estimate how much the variance of a regression coefficient is inflated due to multicollinearity. For each variable in the independent variable selection panel, the corresponding VIF value between the independent and the selected dependent variable is displayed next to the data distribution histogram. A VIF of 1 indicates no correlation between the two variables, while a VIF$>$10 is a strong indication of a potentially severe multicollinearity effect~\cite{vittinghoff2006regression, gareth2013introduction}.
To address such severity, green or yellow backgrounds are used to identify VIF values that are smaller or larger than 10.

\vspace{1.15mm}
\noindent To inspect the details of a feature, the analyst can click on the ``detail button'' 
(\raisebox{0pt}[0pt][0pt]{\raisebox{-0.5ex}{\includegraphics[height=2.5ex]{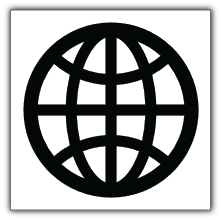}}})
at the end of the corresponding row in the feature list (A2-A4) to open the \emph{detailed feature analysis view}.
This view provides an annotated histogram with details on normality and skewness tests as well as an autogenerated text annotation illustrating the data distribution. To assist analysts, the text also includes interactive suggestions on appropriate transformations to apply to the data. Analysts can click the 
\raisebox{0pt}[0pt][0pt]{\raisebox{-1ex}{\includegraphics[height=3.3ex]{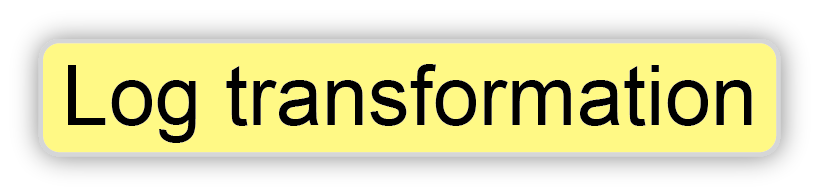}}}
button corresponding to the suggestions to see the transformed distribution.
The Kolmogorov-Smirnov (K-S) test is used to check each feature's normality. When selecting features, the map view updates to show the spatial distribution of the selected variable. 


\vspace{1.4mm} \noindent \textbf{Correlation View:}\;
The correlation view, Figure~\ref{fig:module_configuration} (B), is designed for analyzing relationships between the variables. The view consists of two interactive panels: a scatterplot view (B1) and a scatterplot matrix (B2). Each view contains automatically generated text annotations to highlight bivariate linear correlations based on Pearson correlation tests, which prompts analysts to consider removing redundant features.

\vspace{1.15mm} \noindent The \emph{scatterplot view} shows the correlation between selected variables, Figure~\ref{fig:module_configuration} (B1). Analysts can click on the ``scatterplot button'' (\raisebox{0pt}[0pt][0pt]{\raisebox{-0.8ex}{\includegraphics[height=2.7ex]{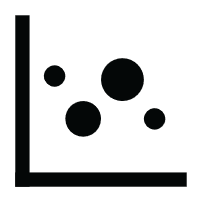}}})
on a row in the feature view to render the corresponding scatterplot view.

\vspace{1.15mm} \noindent The \emph{scatterplot matrix} depicts the correlations among multiple selected independent variables, Figure~\ref{fig:module_configuration} (B2). In addition to the textual suggestions, we annotate the points that show strong linear correlations in \textcolor{orange}{\textbf{orange}} in the scatterplots to help analysts identify potential independent variables.
Analysts can click on these scatterplots to open the detailed distribution information, Figure~\ref{fig:module_configuration} (B3).

\vspace{1.4mm} \noindent \textbf{Map View:}\;
The map view, Figure~\ref{fig:module_configuration} (C), contains two types of choropleth map designs that link with the operations in the feature view and the correlation view. When clicking on a feature's ``detail button'' (\raisebox{0pt}[0pt][0pt]{\raisebox{-0.5ex}{\includegraphics[height=2.5ex]{figures/inline_icon/conf_icon1.png}}}) in the feature view, the \emph{univariate choropleth map} (C1) is rendered to show its spatial distribution with a sequential \textcolor{RoyalBlue}{\textbf{blue}} color scheme in a quantile scale. Clicking on the ``scatterplot button'' (\raisebox{0pt}[0pt][0pt]{\raisebox{-0.8ex}{\includegraphics[height=2.7ex]{figures/inline_icon/conf_icon3.png}}}) on a row in the feature view will update the map to the \emph{bivariate choropleth map} mode (C2) to display the spatial distributions of the dependent variable and the selected independent variable as well as their degree of correlation.
Our bivariate choropleth design is based on Trumbo's diagonal model~\cite{strode2019operationalizing}. As shown in Figure~\ref{fig:module_configuration} (C2), the bivariate choropleth map uses a sequential \textcolor{gray}{\textbf{gray}} color scheme along the main diagonal to show the progression of correlated variables. Combining this with the method from Nusrat et al.~\cite{nusrat2017cartogram}, we use gradients of \textcolor{RoyalBlue}{\textbf{blue}} color (above the diagonal) to depict the regions where the dependent variable values are sufficiently larger than the selected feature, while the gradients of \textcolor{Maroon}{\textbf{red}} color (under the diagonal) are used to represent the regions where the dependent variable values are sufficiently smaller.

All views and text are designed to support novices and experts alike. The text generation templates, designed in conjunction with experts in quantitative geography, provide entry level explanations to guide novices to appropriate parameter selections. Underlying all views are visual enhancements, typically colors, that highlight important features within the data. The underlying combination of text and visual representations serves as a novel guidance mechanism for ESDA.



\subsection{Model Validation}
Once satisfied with model configuration, the analyst can click on the ``Train Model'' button in Figure~\ref{fig:module_configuration} (A1). After the spatial model is trained, the model analysis interface (Figure~\ref{fig:teaser}) is enabled, which equips a suite of views for model validation (Figure~\ref{fig:teaser}(A)), contextualized exploration (Figure~\ref{fig:teaser}(B)) and report authoring (Figure~\ref{fig:teaser}(C)). 

The model validation module (Figure~\ref{fig:teaser} (A)) (\textbf{T2.3}) consists of three views: a model diagnostic panel (A1), a list of local coefficient estimates (A2), and a numerical distribution view (A3). The model diagnostic panel contains detailed information of the global and local diagnostic indicators. Analysts can validate the numerical and spatial distributions of either a local diagnostic indicator or a local coefficient estimate by selecting a row in the corresponding list. By clicking on the ``numerical distribution button'' (\raisebox{0pt}[0pt][0pt]{\raisebox{-0.5ex}{\includegraphics[height=2.5ex]{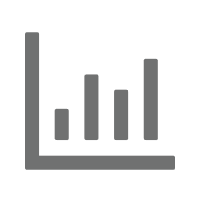}}}) in the list, the histogram of the selected row can be displayed in the numerical distribution view, where the statistical values of the distribution are depicted in the boxplot under the histogram. Meanwhile, the map view shows the coordinated spatial distribution of the selected estimates, Figure~\ref{fig:teaser} (B2). 
The choropleth colors map to a given region's coefficient or diagnostic value. 



We use the corrected Akaike Information Criterion (AICc), the $R^2$, and the adjusted $R^2$ as our global diagnostic indicators, and local $R^2$, Cook's distance, and the standardized residuals as our local indicators to evaluate the overall model performance. Analysts can also explore how the model predicts the relationship between the dependent variable and an independent variable by selecting the corresponding coefficient estimate. The local intercept validates the intrinsic properties of the dependent variable when all the other properties in the model remain constant~\cite{StewartFotheringham2021}. \NEW{To enhance the accessibility of the meaning and context of the above indicators, the framework explains their definitions and the functionalities in the spatial data analysis through pop-over windows. Analysts can click an indicator's name to read the explanation.}

\vspace{1.4mm} \noindent \textbf{Global Diagnostic Indicators:}

\noindent The corrected \textit{Akaike Information Criterion (AICc)} is an estimator of prediction error~\cite{burnham2004multimodel} defined as $AICc=AIC+\frac{2k(k+1)}{n-k-1}$, where $AIC=2k-2logl(\hat{\theta})$, $n$ is the sample size and $k$ is the number of parameters. $AICc$ contains a penalty for model complexity and is a good indicator of model performance when comparing local and global models.

\vspace{1.15mm}
\noindent \textit{$R^2$ and adjusted $R^2$} are used to evaluate how well the model's dependent variable can be explained by the independent variable(s). They represent the proportion of the variance for the dependent variable that is explained by the independent variable(s) in a regression model. 


\vspace{1.4mm}
\noindent \textbf{Local Diagnostic Indicators:}


\noindent The \textit{local $R^2$} indicates how well the dependent variable is explained by the independent variables. For the choropleth map, we encode the \textit{local $R^2$} value for each region using a quantile \textcolor{RoyalBlue}{\textbf{blue}} color scale with single-hue gradients where darker colors indicate higher values.

\vspace{1.15mm}
\noindent \textit{Cook's distance (Cook's $D$)} is used to detect influential outliers in the model. Cook's distance $D_i$ of the observation $i\in\{1,\dots,n\}$ is defined as the summation of all the changes in the regression model when observation (region) $i$ is removed from the model~\cite{cook1977detection}. 
The distribution of the \textit{Cook's $D$} is typically a heavy tail distribution, and we apply a quantile single-hue \textcolor{RoyalBlue}{\textbf{blue}} exponential scale for the choropleth encoding. We only render the regions that are identified as outliers on the map. 

\vspace{1.15mm}
\noindent The \textit{standardized residuals} help identify if the model is well-specified or if key explanatory variables are missing. The residual values represent the over- and underpredictions for each region for the model. We obtain the residual values by subtracting the fitted dependent values from the predicted dependent values. The residual value in overpredicted/underpredicted areas is greater/less than zero respectively. 
For the \textit{standardized residuals}, the over-predicted areas in the choropleth map are shaded using a quantile single-hue \textcolor{RoyalBlue}{\textbf{blue}} color map, and underpredicted areas are shaded using a quantile single-hue \textcolor{Maroon}{\textbf{red}}.

\vspace{1.4mm}
\noindent \textbf{Local Coefficient List:} The local coefficient list contains the estimated local coefficients and local intercepts of the model.

\noindent The \textit{local coefficients} provide information about how each independent variable influences the dependent variable in different regions. In a geographical model, every observation (region) has coefficients ($\beta$) corresponding to each independent variable. The values of these coefficients represent the strength and the type of relationship that the independent variables have with the dependent variable. 


\vspace{1.15mm}
\noindent The \textit{intercepts} represent the expected value for the dependent variable if all the independent variables are zero. Given that all the variables are standardized (0,1) prior to calibration by MGWR, the local intercept estimates indicate the value of $y$ that would be obtained in each location if each location contained a population with average characteristics.

\vspace{1.15mm}
\noindent For the local coefficients and intercept, the framework first performs a t-test to filter out the insignificant coefficients. Then we adopt a quantile single-hue \textcolor{RoyalBlue}{\textbf{blue}} scale to depict the significant positive relationships (positive coefficient values) between the independent variables and dependent variable, and a quantile single-hue \textcolor{Maroon}{\textbf{red}} scale is applied to encode the significant negative relationships (negative coefficient values) in the choropleth map. \textcolor{gray}{\textbf{Gray}} is used to indicate insignificance.

\begin{figure}[t]
    \centering
	\includegraphics[width=\columnwidth]{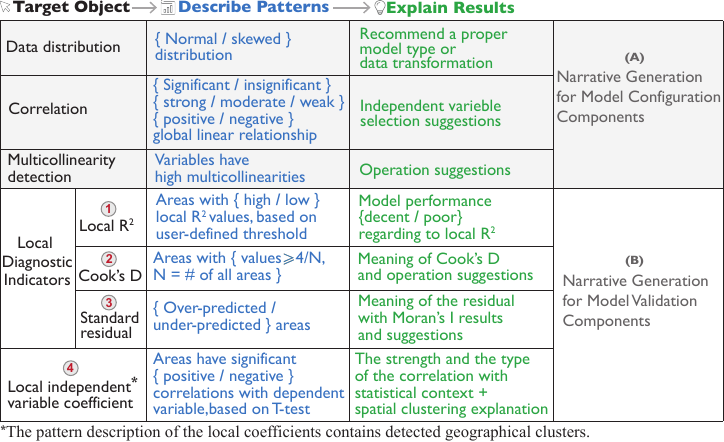}
	\caption{The narrative generation pipeline for the model configuration and model validation components.}
	\label{fig:explanation_templates}
	\vspace{-5mm}
\end{figure}

\subsection{Narrative Generation} 
Underlying the model configuration and validation functions is our narrative generation pipeline.~We build upon previous work from Latif and Beck~\cite{Latif2019} where text template narratives were instantiated to describe bivariate geographic data. However, these templates focused on maximum, minimum, and correlation discussions. 
In this work, we expand their concept to adapt to the generation of explanatory text for the outputs from \NEW{spatial} modeling processes (\textbf{T3.1 - \NEW{T3.3}})\NEW{, and to establish a link to external contextual information (\textbf{T3.4}).} 

\vspace{1.4mm}
\noindent \textbf{\NEW{Template-based Narrative Generation:}} Based on the collaborations with domain experts and the guidelines learned from the geographic data-driven story design practices~\cite{Latif2021}, we design a novel narrative generation pipeline that \NEW{provides template-based explanations for the model configuration and validation functions (Figure~\ref{fig:explanation_templates}). The templates are generated following a three-step process and mainly consist of two phrases: 1) \PHASEONE{the pattern description}, and; 2) \PHASETWO{the result explanation and suggestion containing contextual information}. Figure~A.1 in Appendix A systematically illustrates the narrative generation pipeline and each text template with a corresponding example.}

\vspace{1.15mm}
\noindent \textit{\NEW{Narratives generation for model configuration components:}} \NEW{The \PHASEONE{pattern description phrase} describes the feature analysis results based on specific metrics in different components, while the \PHASETWO{result explanation phrase} generates suggestions on appropriate operations to configure the model or select the parameters (Figure~\ref{fig:explanation_templates} (A)). Two views are dynamically linked with the narrative templates.} In the \NEW{``}detailed feature analysis view\NEW{''} (Figure~\ref{fig:module_configuration} (A5)), a description of the normality test results related to the data distribution of the independent variable is generated.~This is accompanied by recommendations for data transformations. The \NEW{``}correlation view\NEW{''}(Figure~\ref{fig:module_configuration} (B)) \NEW{describes} the correlation test results. This is accompanied by suggestions to prompt analysts to consider removing potentially redundant variables from the list of independent variables in the model.



\begin{figure*}[t!]
    \centering
	\includegraphics[width=1.99\columnwidth]{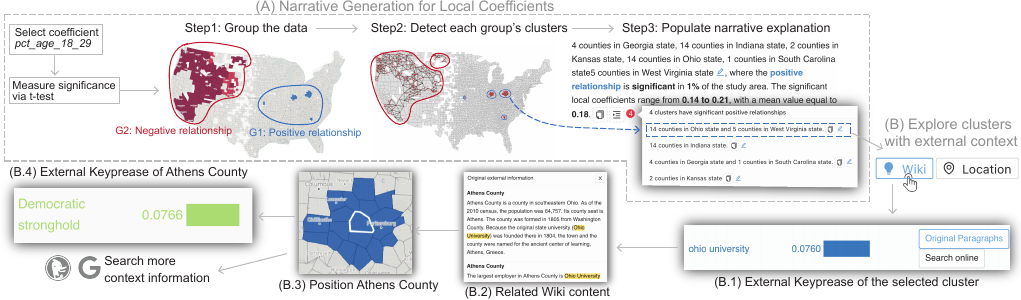}
	\caption{Workflow for exploring the narrative explanations and external context of the local coefficient of the independent variable \textit{pct\_age\_18\_29} (the percentage of the people in the age between 18 and 29) when exploring the MGWR model results for the 2016 U.S. Presidential Election data. (A) shows the narrative generation pipeline for a selected local coefficient across three steps: 1) measures the significance of the coefficient via a t-test and classifies the significant data into two groups, each has a positive/negative relationship with the dependent variable; 2) detects clusters in each group, and; 3) generates a narrative explanation including detected clusters. The analyst finds there is a significant positive inclination for younger voters to vote Democrat in the four detected clusters. Next they explore the external context for a selected cluster in the state of Ohio (B). They find ``Ohio university'' is one of the keyphrases extracted from relevant Wikipedia pages (B.1), then they identify Athens county by browsing the original Wikipedia contents (B.2, B.3). Next, they explore keyphrases related to Athens and realize this county is a democratic stronghold.}
	\label{fig:exploration_case2_workflow}
	\vspace{-4mm}
\end{figure*}


\vspace{1.15mm}
\noindent \textit{\NEW{Narratives generation for model validation components:}} The model validation components contain a suite of interactive views that are closely linked with the map, Figure~\ref{fig:teaser} (B). The \emph{local parameter explanation view}, Figure~\ref{fig:teaser} (B1), interprets model outputs including identified patterns, data distributions, and relevant statistical contexts. The \emph{external information view}, Figure~\ref{fig:teaser} (B3, B4), retrieves external context related to the selected spatial cluster or area in the map. Just as selections in the model's local diagnostic and coefficient list, Figure~\ref{fig:teaser} (A1, A2), populate the map view (Figure~\ref{fig:teaser} (B2)), they also populate the narrative templates in the local parameter explanation view. Analysts can click each parameter's ``show explanation button'' (\raisebox{0pt}[0pt][0pt]{\raisebox{-0.5ex}{\includegraphics[height=2.5ex]{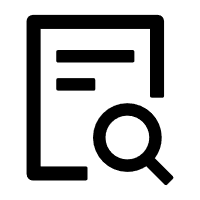}}}) to toggle the explanation panels. \NEW{The local coefficients and three local diagnostic indicators distilled by domain experts that are commonly used in diagnosing local spatial model performance are explained by the template-based narrative generation. The two phrases of the template \PHASEONE{describe the multivariate spatial distributions} and \PHASETWO{incorporate explanations containing contextual information} (Figure~\ref{fig:explanation_templates} (B)).} Here, the spatial distribution is based on the detected geographical patterns and extracted range of values. The generated narrative paragraphs are interactively linked with the map view to enhance comprehension.


\vspace{1.4mm}
\noindent \textbf{Narratives Generation for Spatial Distribution:}
One of the main parts of the text template for describing a selected local parameter is the spatial distribution interpretation. According to the range of values for each local parameter, we classify the parameter data into groups (e.g. the areas that have positive/negative local coefficients). We then populate paragraphs along classified groups to describe the spatial distribution. Each paragraph contains two parts: 1) a \emph{user-definable location identifier} that indicates the spatial locations of the current data group. Analysts are encouraged to use vernacular geographic language (e.g. downtown Chicago, west coast of the U.S.)~\cite{Latif2021} to edit the identified location description to facilitate the communication between the geographic entities and target audience, and; 2) a \emph{narrative description of the classified value range in this location}. We use consistent colors between the map and text to highlight the linkage between spatial narrative and the map view. Hovering the mouse on each paragraph will filter out irrelevant locations on the map.


\vspace{1.4mm}
\noindent \textbf{Local Parameter Explanation View -- Local Diagnostics:} While our narrative generation process follows the same steps and layouts for populating our templates, various model parameters have unique templates to support their explanation, each of which populates the model explanation view. The templates of the model's local diagnostic indicators are introduced below (\textbf{T3.1}):

\vspace{1.15mm}
\noindent \textit{local $R^2$:} The framework generates the spatial distribution narratives by describing the number and location of places with high or low local $R^2$ values, Figure~\ref{fig:explanation_templates} (B.1). Since the validation of local $R^2$ is highly subjective and depends on the specific task~\cite{vittinghoff2006regression}, we use a user-definable threshold to denote high and low local $R^2$ values. We then generate text to explain the meaning and results of different $R^2$ values.



\vspace{1.15mm}
\noindent \textit{Cook's $D$:} We identify outliers and describe their spatial distributions using Cook's $D$ values. We also describe the range, threshold, and meaning of the Cook's $D$, Figure~\ref{fig:explanation_templates} (B.2).

\vspace{1.15mm}
\noindent \textit{Standardized residual:} We describe regions on the map that are over-predicted/under-predicted. The template includes spatial autocorrelation information using Moran's I, Figure~\ref{fig:explanation_templates} (B.3). 

\vspace{1.4mm}
\noindent \textbf{Local Parameter Explanation View -- Local Coefficients:} The local estimates of the coefficients from a local model calibration may contain unique spatial patterns such as outliers and regions with strong spatial autocorrelation. Interpreting these patterns for local coefficients is a key step for understanding local modeling results. In GeoExplainer, underlying the standard narrative generation pipeline in Figure~\ref{fig:explanation_templates} (B.4), we implement an additional set of processes to generate narrative templates to explain the local coefficients and highlight similar patterns (\textbf{T3.2, T3.3}), Figure~\ref{fig:exploration_case2_workflow}.

\vspace{1.15mm}
\noindent \textit{Step1. Group the data:} We group the description of the spatial distributions of the local coefficients with respect to the dependent variable and the selected independent variable into the areas with positive/negative correlations. We describe all the significant coefficients and intercepts, where significance is measured via a t-test that has been adjusted for dependent multiple tests, Figure~\ref{fig:exploration_case2_workflow} (A) - Step 1. 

\vspace{1.15mm}
\noindent \textit{Step2. Detect clusters:} For a selected local coefficient, GeoExplainer automatically detects spatial clusters belonging to each group and generates related interpretations, Figure~\ref{fig:exploration_case2_workflow} (A) - Step 2. Before detection, we derive two queen-adjacency graphs from the significant local coefficients data. One graph contains location nodes with positive coefficient values while the other has nodes with negative coefficient values. For each graph, we perform a community detection algorithm to detect spatial clusters. To balance the time efficiency, community modularity, and the stability of the results~\cite{behera2018performance}, we use the Leiden algorithm~\cite{traag2019louvain} on the server side to identify the spatial communities.

\vspace{1.15mm}
\noindent \textit{Step3. Populate narrative template:} The narrative template is generated in a hierarchical structure with interactions to explain the spatial patterns based on the detected clusters (Figure~\ref{fig:exploration_case2_workflow} (A) - Step3). We populate overview paragraphs that describe the spatial distributions with respect to the positive and negative coefficients first. For a selected coefficient, the overview paragraphs also explain how each independent variable influences the dependent variable in the regions with significant coefficients. For the intercept, we describe how the unmeasurable effects (context) of each region affect the value of the dependent variable. Each overview paragraph will include a toggle list of sub-paragraphs that includes additional information about each cluster detected. The list describes the local geographical patterns along clusters through a user-definable location identifier as described in the spatial distribution templates. Analysts have the flexibility to summarize the patterns with vernacular geographic language. All paragraphs describe the spatial pattern overview and local clusters are interactively linked with the map to call attention to interesting patterns. Analysts can click the paragraph to trigger related external information, Figure~\ref{fig:exploration_case2_workflow} (B).

\vspace{1.4mm}
\noindent \textbf{External Information View:} To help provide further context to the spatial modeling results (\textbf{T3.4}), GeoExplainer links to Wikipedia, enabling quick access to information about local geography, politics, etc. \NEW{Given that distinct datasets exhibit varying spatial resolutions,} we align the Wikipedia information to the dataset's resolution (e.g., county level or community areas for a city). The external information view contains an external keyphrases list (Figure~\ref{fig:teaser} (B3)) of the most important keyphrases for the relevant Wikipedia information. Each keyphrase can populate a list of original Wikipedia content (Figure~\ref{fig:teaser} (B4)) consisting of paragraphs relevant to a particular phrase. Figure~\ref{fig:exploration_case2_workflow} (B.1--B.4) illustrates the external context exploration workflow.

\vspace{1.15mm}
\noindent \textit{External Keyphrases List:} The external keyphrases list displays the top $N$ ($N=20$ by default, can be defined by analysts) most important keyphrases of the Wikipedia information that correspond to a selected range of geographical regions. When analysts explore the narrative explanation of the spatial distributions, they can click the popup ``show keyphrases button'' (\raisebox{0pt}[0pt][0pt]{\raisebox{-0.7ex}{\includegraphics[height=2.3ex]{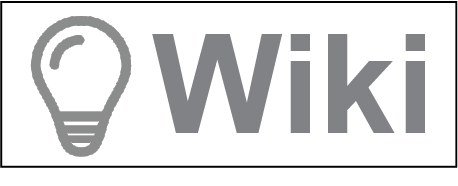}}}) on each paragraph to populate the relevant keyphrases list. Analysts can also trigger the relevant keyphrase list by selecting a range of areas on the map. We extract keyphrases based on a Python implementation of the TextRank algorithm~\cite{PyTextRank}. Each extracted phrase is categorized into one topic that is the same as the corresponding section's topic from Wikipedia. We use a qualitative color scheme
\raisebox{0pt}[0pt][0pt]{\raisebox{-0.5ex}{\includegraphics[height=2.5ex]{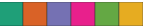}}}
to present those topics. Each row in the keyphrases list contains a keyphrase entity and the rank of this phrase, both are encoded in their topic color. Analysts can click operation buttons on each row to observe the original Wikipedia contents that contain that phrase and search the phrase through the online search engine to gain more context knowledge. By default, we sort the list in descending order based on the ranks of phrases. Analysts can rearrange or filter the list by topics as well.

\vspace{1.15mm}
\noindent \textit{Original Wikipedia Paragraph List:} The original Wikipedia paragraph list displays all the paragraphs that contain the selected keyphrase in the corresponding Wikipedia information. This list can be trigged by clicking
``original paragraphs button'' on each keyphrase row in the external keyphrases list. We highlight the keyphrase with a yellow background in each paragraph. When hovering the mouse over a paragraph, the related map regions are also highlighted.



\subsection{Contextualized Report Authoring}

Throughout the model configuration and validation process, analysts explore model outcomes from a variety of views. To help proceed from data analysis to result communication, GeoExplainer provides a report authoring module that enables the recording of findings and workflow operations, synthesizes explored knowledge, and communicates the results to wider audiences. (Figure~\ref{fig:teaser} (C)) (\textbf{T4.1, T4.2}). 
The module has an expandable report-authoring panel which serves as a collection bin for capturing all the outputs from GeoExplainer. Every text narrative and visualization can be included in this panel. The panel also supports the addition of free-form narratives including adding a new paragraph, editing a paragraph, and deleting a paragraph. The contents in the report-authoring panel are arranged as a sorted list. The order of paragraphs and pictures in the panel can be rearranged by clicking the arrow buttons in the popover toolbox (\raisebox{0pt}[0pt][0pt]{\raisebox{-0.7ex}{\includegraphics[height=2.7ex]{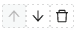}}}) associated with each item in the list. 
The content styles in the narrative paragraphs can be defined by using inline HTML tags with CSS code. For any map added to the report, instead of using the original color scheme, analysts can pick a new color scheme from a list of color schemes. After authoring a report, the framework can export the narrative report to a PDF file automatically by clicking the ``export as PDF button''
(\raisebox{0pt}[0pt][0pt]{\raisebox{-0.9ex}{\includegraphics[height=3.1ex]{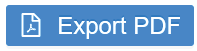}}}).

\subsection{Analytical State Recording}

\NEW{GeoExplainer offers assistance to analysts in recording (\textbf{T5.1}), recovering (\textbf{T5.2}), and sharing (\textbf{T5.3}) their analytical outcomes within the exploration pipeline. At any stage of the pipeline, analysts can save progress into a JSON file with the ``save state'' button. The saved file can be reloaded or shared to reproduce the saved state. Clicking the ``load state'' button retrieves the current state from the saved breakpoint. In addition to the analytical results, training results of the spatial models are automatically recorded once the training process is finished to facilitate model reuse and asynchronous training and analysis processes.}




%% file: content/5_evaluation.tex
\section{Case Study And Expert Interview}
\label{sec:case_study}

In this section, we present a case study and compile results from six expert interviews to demonstrate and evaluate our framework.




\subsection{2016 U.S. Presidential Election Analysis}

In this case study, we invited two domain experts (DE1 and DE2) to analyze the 2016 U.S. presidential election dataset~\cite{StewartFotheringham2021}. The dataset contains 2,813 counties where non-contiguous states and counties with less than 5,000 inhabitants are removed. DE1 is responsible for configuring and training an MGWR model on the election dataset, and DE2 utilizes the explanation and report authoring components.

\vspace{1.4mm} \noindent \textbf{Model Configuration:}
Since the objective of this study is to explore how socioeconomic factors influence voter preferences, DE1 selects \textit{pct\_democrat} (percentage of people who vote for the Democratic Party) as the dependent variable (\textbf{T2.1}) from the Original Feature List (Figure~\ref{fig:module_configuration} (A4)) after he loads the data. He then browses the populated univariate choropleth map (Figure~\ref{fig:module_configuration} (C1)) and follows the operation recommendation to normalize the dependent variable by clicking the ``log transformation'' button (\textbf{T2.2}), Figure~\ref{fig:module_configuration} (A5).

Next, DE1 needs to choose independent variables for the model (\textbf{T2.1}). By dragging and dropping interested features to the Independent Variable List (Figure~\ref{fig:module_configuration} (A3)), DE1 explores the correlation between each independent variable candidate and the dependent variable. He observes a strong negative global relationship between \textit{pct\_gop} (percentage of people who vote for the Republican Party) and \textit{pct\_democrat} in the Correlation View, Figure~\ref{fig:module_configuration} (B1). 
Next, DE1 explores the bi-variate map and notes that urban and coastal areas are more likely to vote for the Democratic Party, indicating that there may be underlying latent factors that determine voter preferences, Figure~\ref{fig:module_configuration} (C2). DE1 notices high multicollinearity among many socioeconomic factors according to the VIF scores and scatterplot matrix, Figure~\ref{fig:module_configuration} (B2, B3), after discussion with DE2, DE1 compares the trained model's global $R^2$ and AICc values (\textbf{T2.3}) by trying the different combinations of the socioeconomic factors. Based on the exploration and discussion, DE1 chooses the following independent variables: [\textit{sex\_ratio}, \textit{pct\_black}, \textit{pct\_hisp} (percentage of Hispanic population), \textit{pct\_bach} (\% of people with a bachelor's degree), \textit{income}, \textit{pct\_65\_over}, \textit{pct\_age\_18\_29}, \textit{gini}, \textit{pct\_manuf} (manufacture), \textit{log\_pop\_den} (log of population density, a measure of 'urbanness'), \textit{pct\_3rd\_party} (votes for a third-party candidate), \textit{turn\_out} (voter turnout), \textit{pct\_FB} (foreign-born), \textit{pct\_insured} (people with health insurance)]. DE1 then trains the model and validates that the model's local performance is reasonable from the generated explanation of local residuals (\textbf{T3.1}). DE1 saves the analytical state (\textbf{T5.1}) and shares the state with DE2 (\textbf{T5.3}).

\vspace{1.4mm} \noindent \textbf{Exploring the Model Coefficients\NEW{:}}
After loading the analytical state file (\textbf{T5.2}), DE2 explores the local coefficient list and notices an unusual spatial distribution of the coefficient \textit{pct\_age\_18\_29} (the percentage of the people in the age between 18 and 29) in the map, Figure~\ref{fig:teaser} (B2). DE2 reads the narrative explanations in the Parameter Explanation View (\textbf{T3.1}) and realizes that the vast majority of county-specific local coefficients of \textit{pct\_age\_18\_29} are insignificant, except for counties in Southern California and the counties throughout much of the Northwest. The choropleth map shows younger voters (aged eighteen to twenty-nine) have a significant inclination to vote Republican in these regions. DE2 also notices that the framework has detected two small clusters in Indiana and Ohio (\textbf{T3.2}), where there is a significant positive inclination for younger voters to vote Democrat (\textbf{T3.3}). DE2 then explores relevant context information populated in the External Keyphrases List (\textbf{T3.4}). While DE2 is familiar with the region in Indiana, quickly identifying it as the home of Indiana University, the region in Ohio is ambiguous. DE2 observed ``ohio university'' as one of the high-ranking keyphrases of this cluster, and realized Athens County, Ohio in the cluster is a Democratic stronghold and home to Ohio University in the corresponding Wikipedia Paragraph List. Figure~\ref{fig:exploration_case2_workflow} illustrates the workflow of exploring the narrative explanation and external context. Here DE2 noted the seamless ability of GeoExplainer to pull contextual information about unfamiliar regions to help hypothesis development.

After recording these findings and relevant maps in the report authoring panel (\textbf{T4.1}), DE2 
explores the local estimates of the intercept. In MGWR, the local estimates of the intercept indicate the impact of geographical context on the dependent variable while holding everything else in the model constant~\cite{StewartFotheringham2021}.
\begin{wrapfigure}{r}{0.17\textwidth}
  \vspace{-20pt}
  \begin{center}
  \hspace{-10pt}
    \includegraphics[width=0.17\textwidth]{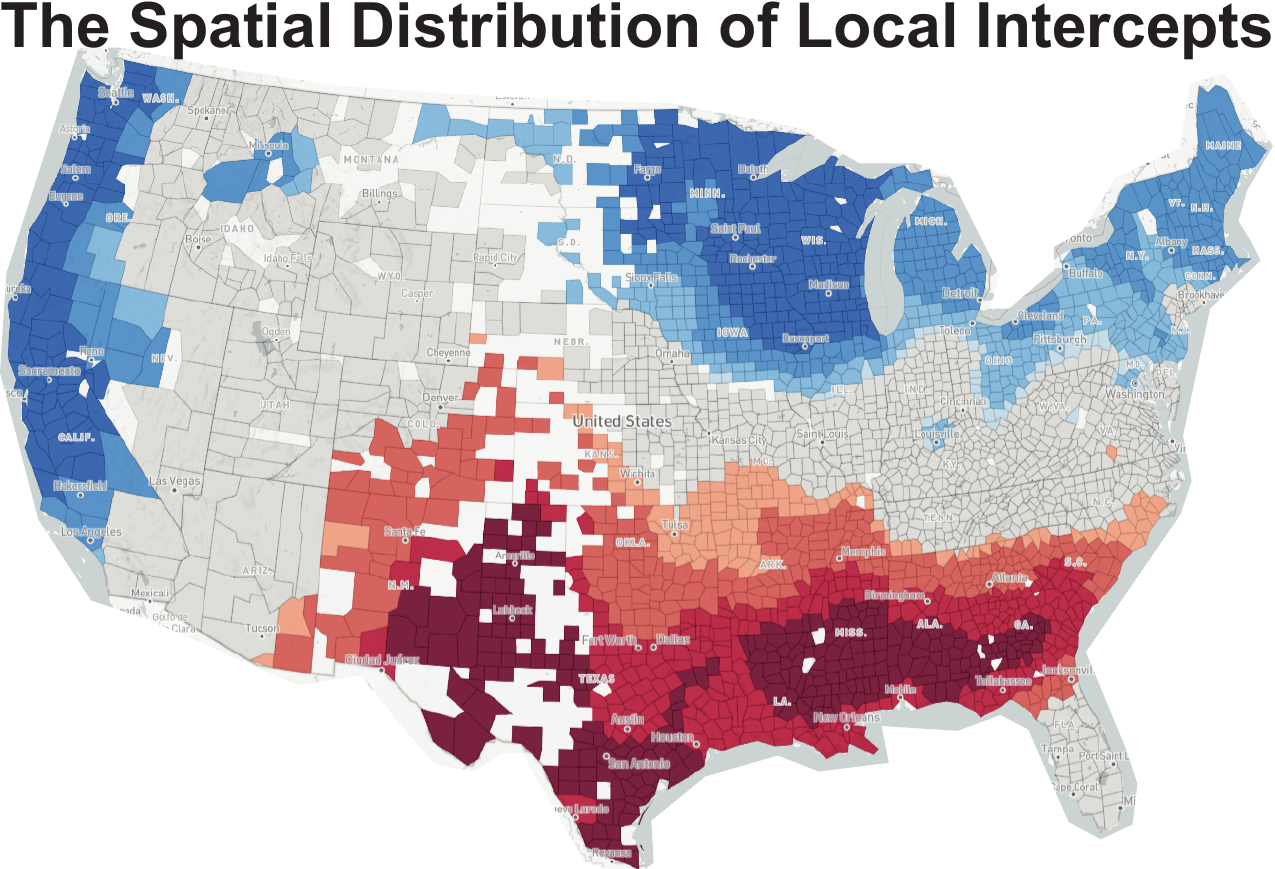}
  \end{center}
  \vspace{-20pt}
\end{wrapfigure}
In this case, the local intercepts indicate the intrinsic support for the Democratic Party or the Republican Party if all counties had exactly the same mix of population. DE2 explores the spatial distribution of the local intercepts on the map (\textbf{T2.3}) and the Parameter Explanation View (\textbf{T3.1}). Here, the spatial distribution and the relevant narrative interpretation suggest that the counties throughout New England, the upper Midwest, and down the Pacific Coast intrinsically tend to vote Democrat, and counties throughout the South, except for Florida, are intrinsically more likely to vote Republican. This spatial distribution pattern is particularly informative to the expert because it represents the unseen, but important, influence of geographical context on behavior. It also represents how each county would vote if it contained an average population composition.
That is, if all counties across the US had exactly the same population composition, the distribution of voting behavior would not be constant but as shown on the map. The expert adds this map and the corresponding interpretation to the report panel (\textbf{T4.1}). 

\vspace{1.4mm} \noindent \textbf{\NEW{Performance:}} \NEW{Visual elements, including charts, maps, and narrative generation, load and display at a sub-second rate. Spatial cluster detection and keyphrase extraction take roughly 2 seconds, respectively. The MGWR training process took a duration of around 186 seconds. The scalability of the framework is discussed in Sect.6.}

\subsection{Expert Interviews}
\NEW{GeoExplainer is designed for spatial data analysts who use geographic models to analyze spatial-related phenomena.} To further assess our framework, we conducted a group interview with six \NEW{\sout{collaborators}}\NEW{participants} (E0 to E5) via video-conferences. \NEW{Three participants are geographers who have expertise in spatial data analysis for at least 5 years. The other three participants are Ph.D. students in geographic analysis. All participants have experience using spatial regression models.} We recorded the participants’ audio and computer screen for subsequent analysis with their consent.
We first introduced the background and the analytical tasks of our work, followed by a detailed explanation of the analytical workflow and the interface with a case study. Then, the experts were encouraged to explore the dataset and develop a report using GeoExplainer. During the interview, participants were asked to share their screens with us; they were encouraged to ask questions when necessary. The interview lasted approximately 90 minutes. At the end of the interview, we collected feedback and distributed two questionnaires (a usability evaluation and a functionality questionnaire) to each of the domain experts. Both questionnaires include a series of 7-point Likert-scale questions and are provided in the supplementary material. Figure~\ref{fig:evaluation_result} shows the aggregate results of the surveys.

\begin{figure}[t!]
  \centering
\includegraphics[width=\columnwidth]{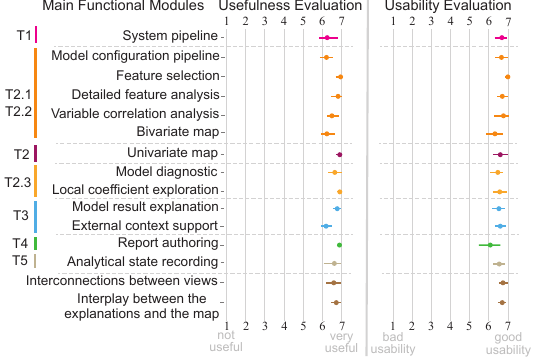}
\caption{Average ratings from six participants regarding the usability and usefulness of the \NEW{main functional modules and interactions} of GeoExplainer. Modules are grouped based on the analytical tasks.}
\label{fig:evaluation_result}
\vspace{-5mm}
\end{figure}


Combining the questionnaire scores and the feedback from experts, our \textit{\textbf{unified analytical pipeline}} received a high average score (6.23/7) with the respect to its perceived usefulness. The experts commented that the integration of the exploratory spatial data analysis (ESDA) process with the narrative interpretations (\textbf{T1.1}) can improve the efficiency of their spatial analytical tasks and save them a lot of time compared to the traditional tools. E1 noted that \textit{``I am happy that the framework has already established a good analytical pipeline, now I can review my analysis results more efficiently without bothering to import different packages to build my own workflow.''} 
The average score (6.76/7) of the perceived ease of use from the questionnaire further indicates that the dynamic narrative visualization design in the framework is able to promote usability. E1 appreciated the flexibility of the interactions and the intuitive visual design of the multiple modules' layout. E2 also had a positive comment with regard to the interactive visualization widgets commenting that \textit{``Having used other drag-and-drop type visualization software such as Tableau and ArcGIS, I found the interaction of GeoExplainer to be really straightforward and easy to use even without the tutorial video.''} E3 further added that they were impressed by the single dashboard design for the model validation and reporting features. E4 and E5 both like the smooth transitions between different views. 

Another favorite feature of GeoExplainer was the \textit{\textbf{narrative explanations and their interplay with the map}}. All narrative explanation views and inline annotation modules received high average ratings regarding their usefulness and usability (Figure~\ref{fig:evaluation_result} (T3)). \NEW{The participants agreed that the narrative explanations generated were deemed satisfactory and effective for aiding their analytical tasks. E1 showed a preference for concise interpretation paragraphs that interactively filter the relevant map layer while examining coefficients and local diagnostics (\textbf{T3.1, T3.3}). He mentioned that these explanations facilitated a rapid comprehension of local spatial distributions and guided his focus towards relevant areas on the map. E4 and E5 both appreciated the hierarchical interpretations of spatial clusters for local coefficients (\textbf{T3.2}). They noted that the local context information for each spatial cluster was readily accessible and incorporated into the report. They also appreciated the Wikipedia keyphrases lookup feature with supplemental context information for the selected geographic cluster (\textbf{T3.4}). All student participants found the explanations of the statistical indicators to be highly valuable, as they facilitated the evaluation and understanding of the model's performance and saved considerable time and effort.}

The \textit{\textbf{report authoring and the analytical state record functions}} both received high ratings for their usefulness and utility. 
E2 mentioned: \textit{``I can add whatever worth noticing to the report when exploring [...] I also like the flexibility to arrange my findings and edit them into a report.''} E1 and E3 appreciated they can synthesize their findings into a communicable report without using any 3rd party tools (\textbf{T4.2}). E5 especially mentioned analytical process recording function is very useful when she needs to gradually analyze a model's results or share the idea with other collaborators (\textbf{T5}). E5 said: \textit{``Now I can save my analytics state, do other jobs, and back to the analysis days later without the need to retrain the model. [...] I also invited my colleagues to do their analysis based on my configured model on GeoExplainer ...''}

\vspace{1.4mm} \noindent \textbf{Limitations:} The domain experts also identified some limitations of our current framework. E0 noted that the map layer loading time needs to be further optimized, and E0, E1, and E2 expected the map visualization in our framework to support more user-defined operations, such as the selection of color and filtering rules. E2 suggested that the framework's explanatory features could be further expanded to include an explanation of the traditional ordinary least squares (OLS) model's result. E4 suggested that we should work to provide more content automation from the external search function to the model results.

%% file: content/6_conclusion.tex
\section{Discussion and Conclusions}
\label{sec:conclusion}

In this paper, we propose a visual analytics framework, GeoExplainer, for supporting spatial data modeling, contextualization, and reporting. We introduce a suite of interactive views that are supported by a template-based narrative generator to provide explanatory text that can directly be utilized in report generation. As analysts create their spatial models, our framework automatically highlights potential issues in model parameter selections, utilizes template-based text generation to summarize model outputs in detail, and links with external knowledge repositories, such as Wikipedia, to provide interactive annotations.

The innovations of this system are in the use of text generation for multivariate spatial models that require locational information to support answering ``why'' questions about spatial phenomena. While the applied visualizations utilize well-known components, the integration of automatic analysis and human-in-the-loop for exploring and contextualizing model results goes well beyond current systems that support general descriptive statistics (mean, median, correlation). Contributions include methods to automatically detect and explain spatial patterns, all while enabling analysts to capture and share their workflow. The proposed text templates go beyond simple descriptive statistics (mean, mode, correlation) to focus on multivariate patterns and spatial phenomena. As models become more complex, such advances help bridge the gap for explainable artificial intelligence.

\vspace{1.4mm} \noindent \textbf{Generalizability:} \NEW{Our framework is designed support different spatial regression models that shares the same input and output protocol. Owing to the modularized design, new models can be swapped in to replace MGWR adopted in the case study. The asynchronous model training process can further enable the integration of models with varying computational complexities. For enhancing explanations of diverse spatial models, our narrative generation function, informed by domain experts, selectively presents and elucidates three commonly-used diagnostic indicators (local $R^2$, Cook's distance, and standard residual) for local spatial model validation.} The input of our framework can be an arbitrary spatial dataset where every observation in the dataset has features corresponding to a geographic location. With respect to the data format, GeoJSON is used in the current prototype, but additional geographic file types can also be supported, such as ESRI shapefiles. 

\vspace{1.4mm} \noindent \textbf{\NEW{Scalability:}} \NEW{Advanced spatial models, e.g., MGWR, may be resource-intensive as the data scale expands~\cite{doi:10.1080/13658816.2020.1720692}. GeoExplainer adapts to varying data scales with the aid of an asynchronized state recording functionality, as described in Sect.4.5. To optimize the loading duration of the views for large datasets, data to be rendered in the visual interface can be processed concurrently with the model training process.}

\vspace{1.4mm} \noindent \textbf{\NEW{Target Audience:}} \NEW{As mentioned in Sect.4, our target audience comprises spatial data analysts requiring enhanced comprehension and communication of model results. Some of the features in our framework like interactive explanations of statistical indicators can also be used by novice audiences with limited spatial analysis knowledge. However, the framework needs further improvement and evaluation of the functionality for novices.}

\vspace{1.4mm} \noindent \textbf{Limitations and Future Work:} \NEW{While template-based narrative explanations facilitate quality control and maintain consistency, these templates are task-oriented and necessitate fine-tuning by domain experts. Possible extensions grounded in large language models may enhance flexibility and adaptability of the narrative generation process. In terms of extending the framework's applicability to a general audience, there are current plans to deploy GeoExplainer in large classroom settings to further assess the report authoring module for junior analysts. We also expect to improve the contextualization support by incorporating knowledge graph-based extraction techniques.}




%% file: content/supplemental_materials.tex
\section*{Supplemental Materials}
\label{sec:supplemental_materials}

The supplemental materials include 
(1) a PDF example report generated by the case study, (2) a JSON file as an example state-saving file of the case study in Section 5.1, (3) example expert interview questions in a PDF file, (4) an Excel file containing the evaluation results for creating Figure~\ref{fig:evaluation_result}, (5) an appendix, and (6) a demo video.
All code is available at \url{https://github.com/VADERASU/geoexplainer}.


%% file: content/7_appendix.tex
\clearpage
\begin{landscape}
\section{Narrative Generation Templates}
\label{appendix:narrative_templates}

\renewcommand\thefigure{\thesection.\arabic{figure}}  
\setcounter{figure}{0}    

\begin{figure}[h]
    \centering
	\includegraphics[width=1.2\textwidth]{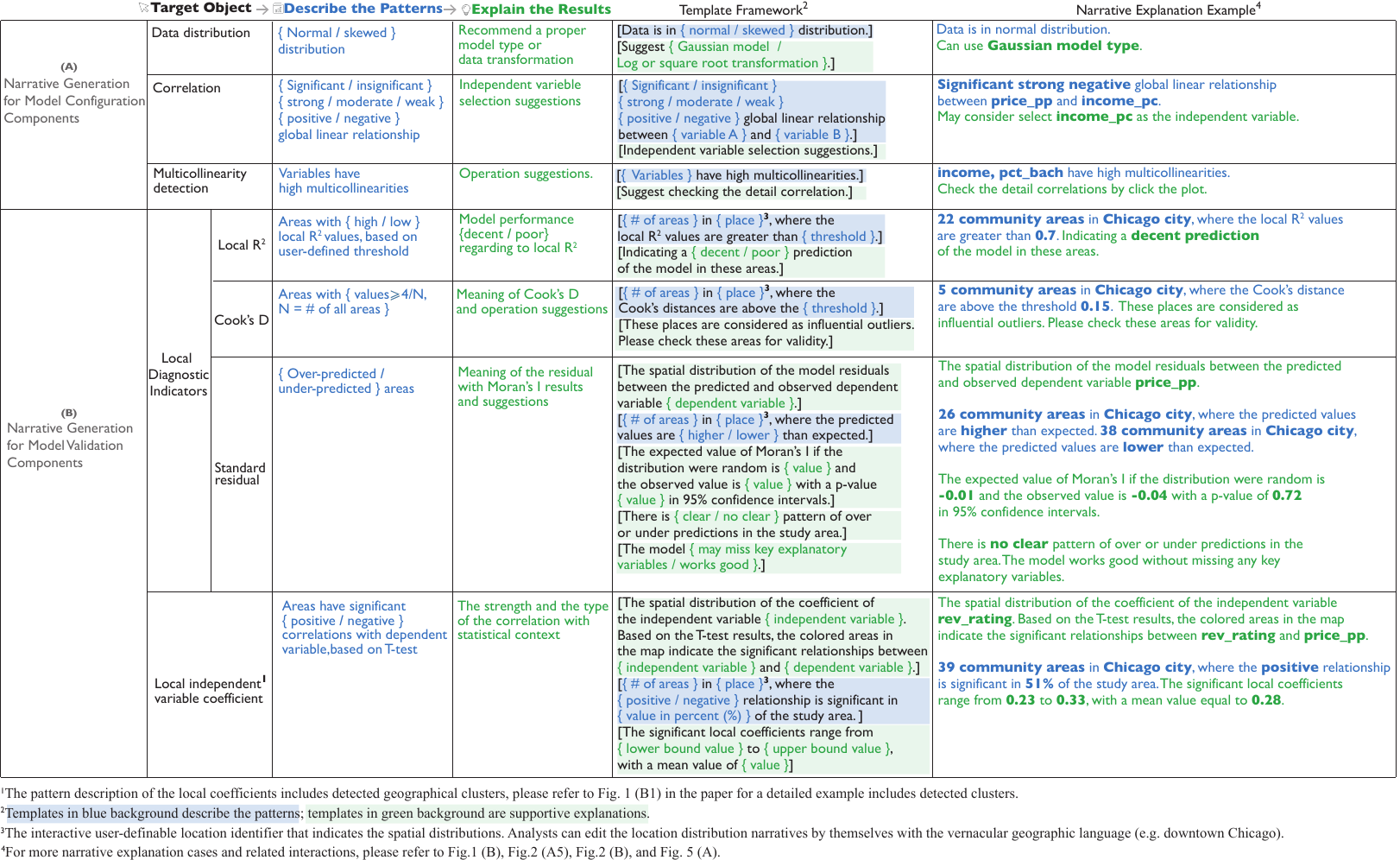}
	\caption{The narrative generation pipeline and text templates with examples.}
	\label{fig:full_text_template}
	\vspace{-4mm}
\end{figure}
\end{landscape}